\documentclass[prd,aps,twocolumn,nofootinbib]{revtex4}%,%onecolumn,%groupedaddress]{revtex4}
\usepackage{epsfig}
\usepackage{amsfonts}
\usepackage{rotating}
\usepackage{sparticles}
\usepackage{dcolumn}
\usepackage{bm}
\usepackage{hyperref}
\hypersetup{
    bookmarks=true,         % show bookmarks bar?
    unicode=false,          % non-Latin characters in Acrobat’s bookmarks
    pdftoolbar=true,        % show Acrobat’s toolbar?
    pdfmenubar=true,        % show Acrobat’s menu?
    pdffitwindow=false,     % window fit to page when opened
    pdfstartview={FitH},    % fits the width of the page to the window
    pdftitle={My title},    % title
    pdfauthor={Author},     % author
    pdfsubject={Subject},   % subject of the document
    pdfcreator={Creator},   % creator of the document
    pdfproducer={Producer}, % producer of the document
    pdfkeywords={keyword1} {key2} {key3}, % list of keywords
    pdfnewwindow=true,      % links in new window
    colorlinks=true,       % false: boxed links; true: colored links
    linkcolor=red,          % color of internal links
    citecolor=cyan,        % color of links to bibliography
    filecolor=magenta,      % color of file links
    urlcolor=green,           % color of external links
    linktocpage=true
}
\usepackage{amsmath,amssymb,amsthm,graphicx,latexsym}
\usepackage{subfigure}
\usepackage{pstricks}
\usepackage{color}

\usepackage{mathrsfs}

\newcommand{\be}{\begin{equation}}
\newcommand{\ee}{\end{equation}}
\newcommand{\bea}{\begin{eqnarray}}
\newcommand{\eea}{\end{eqnarray}}
\newcommand{\bml}{\begin{subequations}}
\newcommand{\eml}{\end{subequations}}
\newcommand{\bfig}{\begin{figure}}
\newcommand{\efig}{\end{figure}}

\begin{document}

\title{%A precise constraint on inflection-point model of inflationary magnetogenesis via various cosmoparticle probes
Inflamagnetogenesis redux: Unzipping sub-Planckian inflation via various cosmoparticle probes}

\author{Sayantan Choudhury$^{}$\footnote{Electronic address: {sayanphysicsisi@gmail.com}} ${}^{}$
}
\affiliation{Physics and Applied Mathematics Unit, Indian Statistical Institute, 203 B.T. Road, Kolkata 700 108, INDIA
}

%\vspace{5ex}
%\date{\today}
\begin{abstract}
In this paper I introduce a precise constraint on primordial magnetogenesis,
 for a generic
class of single-field inflationary model followed by small field excursion below the Planck scale.
I also establish a connection between the magnetic field at the
present epoch and primordial gravity waves ($r$) via non-vanishing CP asymmetry parameter ($\epsilon_{\bf CP}$), which triggers the leptogenesis scenario.  
Finally, I explore various hidden cosmophenomenological features of 
theoretical CMB B-mode polarization spectra, which can be treated as a significant probe
to put further stringent constraint on low and high scale small field inflationary models after releasing the Planck B-mode polarization data.
\end{abstract}

\maketitle
%%%%%%%%%%%%%%%%%%%%%%%%%%%%%%%%%%%%%%%%%%%%%%%%%%%%%%%%%%%%%%%%%%%%%%%%%%%%%%%%%%%%%%%%%%%%%%%%%%%%%%%%%%%%%%%%%%%%%%%%%%%%%%%%%%%%%%%%%%%%%%%%%%%%%%%%%%%%%%%%%%%%%%%%%%%%%%%%%%%%%%%%%%%%%%%%%%%%%%%%%%%%%%%%%%%%%%%%%%%%%%%%%%%%%%%%%%%%%%
%%%%%%%%%%%%%%%%%%%%%%%%%%%%%%%%%%%%%%%%%%%%%%%%%%%%%%%%%%%%%%%%%%%%%%%%%%%%%%%%%%%%%%%%%%%%%%%%%%%%%%%%%%%%%%%%%%%%%%%%%%%%%%%%%%%%%%%%%%%%%%%%%%%%%%%%%%%%%%%%%%%%%%%%%%%%%%%%%%%%%%%%%%%%%%%%%%%%%%%%%%%%%%%%%%%%%%%%%%%%%%%%%%%%%%%%%%%%%%%%%%%%%%%%%%%%%%
 
In principle the CMB
 polarization field is conventionally decomposed into a sum of parity-even curl-free ``E-mode'' and parity-odd gradient free ``B-mode''
 contributions \cite{Kamionkowski:1996ks}; since the B-mode polarization is absent from linear order scalar density perturbations which are the
dominant source of CMB temperature and E-mode anisotropies,
serves as a powerful test of various hidden aspects of cosmological physics, including primordial tensor and vector perturbations.
The other complementary origin of the B-type polarization is Faraday rotation of the orientation of linear polarization due to the
presence of a primordial magnetic field. In such a physical scenario, if a purely ``E-mode'' polarization pattern is rotated by an angle, $\theta_{F}=\pi/4$, then it
transforms into a purely ``B-mode'' polarization and for $\theta_{F}<<\pi/4$, it will generate a component of the total B-type polarization.
 Besides the Faraday rotation effect also induces non-zero parity-odd cross correlations of type
TB and EB, which are absent for standard cosmological scenarios. Primordial magnetic fields are
interesting topic in the present day research in ``particle cosmology'' due to three prime reasons:$\Rightarrow${\bf (1)} they could serve as seed for the observed magnetic fields in galaxies and galaxy clusters,
$\Rightarrow${\bf (2)} they can be treated as the
clean probe of signal of primordial
tensor perturbations, the detection of which would directly constrain the
scale of inflation and $\Rightarrow${\bf (3)} also act as a probe of gravitational lensing.

The earlier results obtained from the various famous CMB experiments by Planck, WMAP9 along
 with various combined constraints \cite{Ade:2013uln,Ade:2013zuv,Hinshaw:2012aka} have placed upper limits 
either on the B-mode polarization anisotropy or on the tensor-to-scalar ratio obtained from the inflationary picture. Very recently SPT confirms 
the first detection signatures of B modes sourced by
gravitational lensing \cite{Hanson:2013hsb}. In CMB the B-modes is made up of primordial gravity waves \cite{Seljak:1996gy} ,
 non-Gaussian fluctuations in the primordial fluctuations \cite{Zaldarriaga:1998ar} and the
contribution from the gravitational lensing. Recently BICEP2 \cite{Ade:2014xna} also detects the B-modes which confirm the signature of the primordial gravity waves via tensor-to-scalar 
ratio within the window $0.15<r_{\star}<0.27$ (here $\star$ represents momentum pivot scale), as it peaks around the 
multipole $l\sim 80$ in the CMB BB angular power spectra~\footnote{For any required multipole $l_{reqd}$, the corresponding value of the momentum scale can be calculated by using the relation
$k_{reqd}=\frac{l_{reqd}}{\eta_{0}\pi}$, where $\eta_{0}\sim 14000$ {\rm Mpc} be the conformal time at the present epoch.}. However, to make sure 
further about this crucial issue we should need to separate the contribution of the primordial non-Gaussianity and other significant components as well as from the observed B modes from BICEP2.
 However as the amount of non-Gaussian fluctuations are very small as predicted 
by the recent observation from Planck \cite{Ade:2013ydc}, I, henceforth, neglect such contribution from my theoretical analysis. 

\begin{figure}[htb]
\centering
\includegraphics[width=9cm,height=8.8cm]{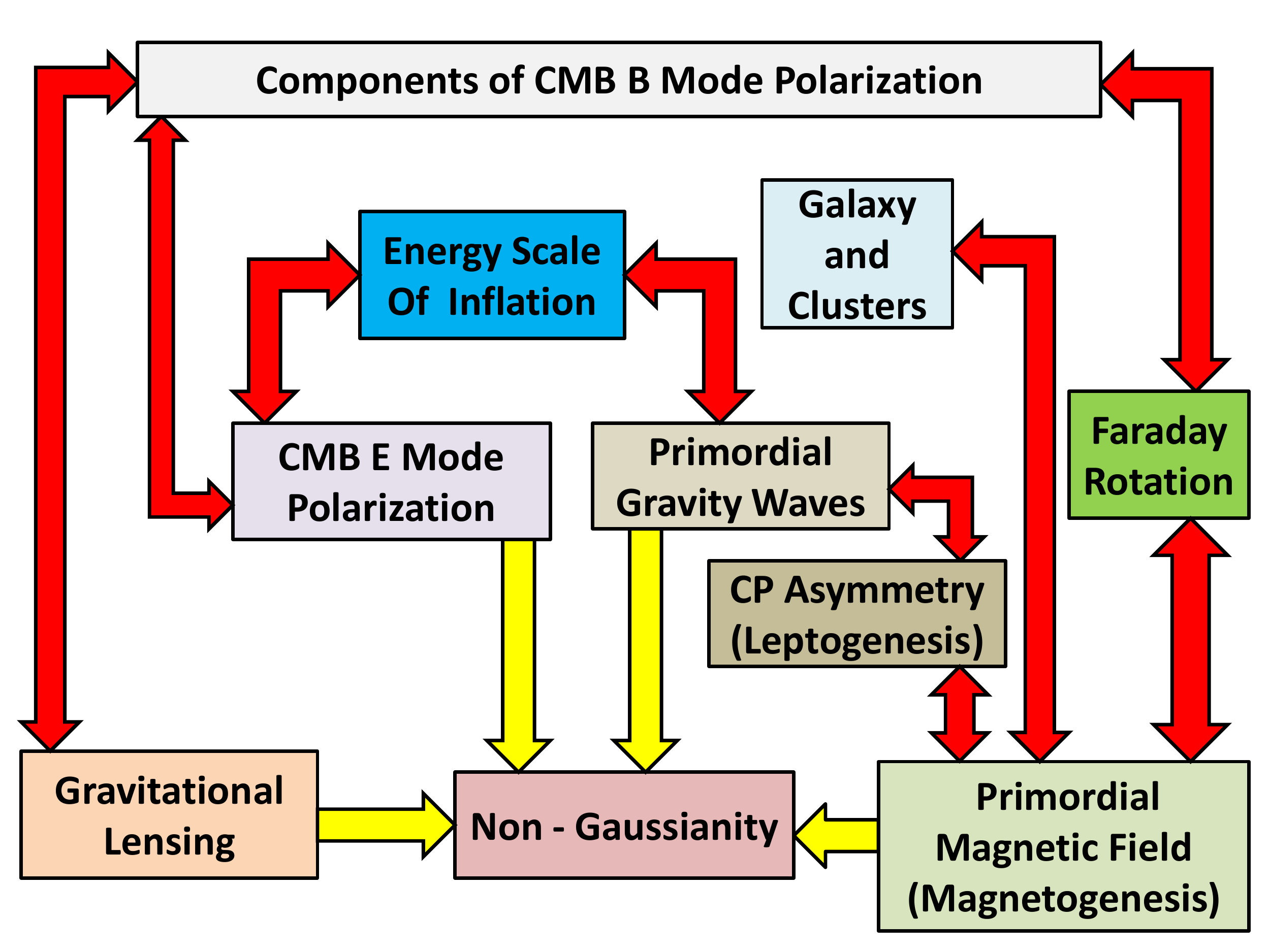}
\caption{Schematic diagram of the components of CMB B mode polarization.
}
\label{f1aa}
\end{figure}

In Fig~(\ref{f1aa}) I have explicitly shown the schematic representation of the significant components of the CMB B mode polarization.
In the most generalized physical prescription the CP asymmetry parameter, primordial gravity waves and Galaxy-clusters play crucial role in CMB BB correlation.
Here CP asymmetry parameter is one of the source for generating primordial magnetic field from particle physics which triggers the role of leptogenesis
 scenario in the present context. Additionally it is important to note that the magnetic
 fields appear to be present in all galaxies and galaxy clusters \cite{Naoz:2013wla}. Recent measurements
indicate that a weak magnetic field may be present even in the smooth low density intergalactic
medium. One explanation for these observations is that a seed magnetic field was generated by some
unknown mechanism in early Universe, and was later amplified by various dynamos
in nonlinear objects like galaxies and clusters. In such a case primordial magnetic field is expected
to be generated in the early Universe on purely linear scales through vorticity induced by scale-
dependent temperature fluctuations or equivalently, a spatially varying speed of sound of the gas.
Last but not the least primordial gravity waves originating from inflationary paradigm via tensor-to-scalar ratio is one of the prime component
which makes a connection between the inflationary magnetogenesis and leptogenesis scenario. Now it is paramount to note that if inflation can be embedded in 
effective field theory setup guided by particle physics framework in which the Ultra Violet (UV) cut-off of the scale of the gravity, $\Lambda_{\bf UV}\leq M_{p}$
~\footnote{Here $M_{p}=2.43\times 10^{18}~{\rm GeV}$ be the reduced Planck mass.}, then in such phenomenological setup it is possible to generate required amount of 
primordial magnetic field
from inflationary paradigm as well as from leptogenesis via CP asymmetry parameter. Also within this particle phenomenological 
framework it is possible to make a connecting relation between both the sectors. As for a specific case if the inflationary model is fully embedded within
a particle theory such as that of gauge invariant supersymmetric flat
directions of Minimal Supersymmetric Standard Model
(MSSM),  MSSM$\otimes U(1)_{B-L}$, Next to Minimal Supersymmetric Standard Model
(NMSSM) etc. 
 In Fig~(\ref{f1aa}) we have used two types of arrow-bidirectional and unidirectional. The bidirectional arrows are applicable in such cases where the reconstruction
techniques and de-lensing methodologies are applicable. On the other hand, unidirectional arrows are explicitly used for those physical situations where such reverse techniques
are not at all applicable. However, it is a very challenging task in particle cosmology to materialize the reconstruction
techniques and de-lensing methodologies very smoothly~\footnote{See refs.\cite{Choudhury:2014kma,Sigurdson:2005cp,Hunt:2013bha,Guo:2011re,Hu:2014aua}
 where the technical issues of these aspects are discussed in details.}.
   
\begin{figure}[htb]
\centering
\includegraphics[width=9cm,height=8.8cm]{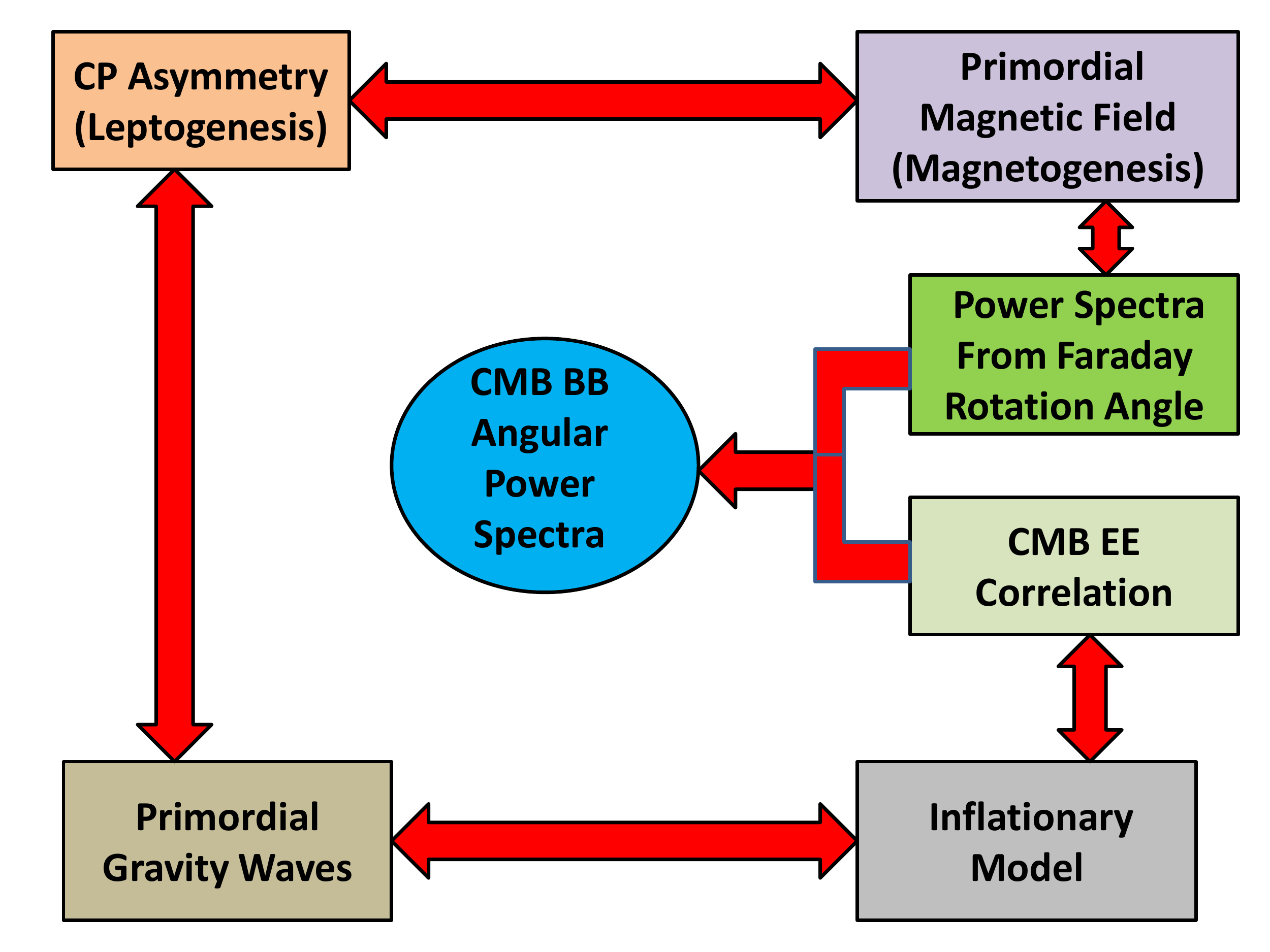}
\caption{Schematic representation of the proposed cosmophenomenological technique to determine theoretical CMB B-mode polarization spectra from BB correlation
using various cosmoparticle probes obtained from the low and high scale sub-Planckian models of inflationary setup.
}
\label{f1a}
\end{figure}

Further in Fig~(\ref{f1a}) I have depicted the
proposed cosmophenomenological technique
 to determine theoretical CMB B-mode polarization spectra from BB correlation
using the models of inflationary setup where the scale of inflation, field excursion ($\Delta\phi=|\phi_{cmb}-\phi_{end}|$
~\footnote{Here $\phi_{cmb}$ and $\phi_{end}$ are the field values at the horizon crossing and at the end of inflation respectively.})
 and VEV ($\phi_0$) are lying below the Planck scale. Fig~(\ref{f1a}) implies that once we know a specific model of sub-Planckian inflation, it is possible to 
calculate the contribution for primordial gravitational waves via tensor-to-scalar ratio and further using this input, it is possible to predict the contribution of 
CP asymmetry which will participate in leptogenesis scenario. Hence, using this information, the amount of primordial magnetic field can be estimated from a given 
sub-Planckian inflationary model which will further generate the CMB B mode polarization spectra. 

%%%%%%%%%%%%%%%%%%FIGURE%%%%%%%%%%%%%%%%%%%%%%%%%%%%%%%%%

%%%%%%%%%%%%%%%%%%%%%%%%%%%%%%%%%%%%%%%%%%%%%%%%%%%%%%%%%%%%%%%%%%%%%%%%%%%%%%%
%%%%%%%%%%%%%%%%%%%%%%%%%%%%%%%%%%%%%%%%%%%%%%%%%%%%%%%%%%%%%%%%%%%%%%%%%%%%%%%%%
The aim of this paper
is to present a precise constraint on primordial magnetogenesis
 for a generic
class of single-field inflection point inflationary model followed by small
 field excursion characterized by two features 
\cite{Allahverdi:2006we,Choudhury:2013iaa,Choudhury:2013woa,Choudhury:2013jya,Choudhury:2011jt,Choudhury:2014kma,Choudhury:2014wsa}
~\footnote{In this paper as my prime motivation to embed the particle theory within effective field theory setup I only concentrate on the small field models of inflation.
But once I move to the large field models of inflation which are guided by super-Planckian VEV and field excursion, the identification of the magnetic field
obtained from leptogenesis and inflation is not possible as the effective field theory prescription does not hold in such situations.}, $\Rightarrow${\bf (1)} $\phi_0\leq M_p$ - vev of the inflaton must be bounded by the cut-off of the particle theory
and $\Rightarrow${\bf (2)} $\Delta\phi\leq M_p$- the inflaton potential has to be flat enough  during which a successful
inflation can occur. Using this sub-Planckian small field excursion criteria in this paper I first establish a connection between the magnetic field at the
present epoch and primordial gravity waves ($r$) via non-vanishing CP asymmetry parameter ($\epsilon_{\bf CP}$)  
from which I further explore various hidden cosmophenomenological features of 
CMB B-mode polarization spectra which can be useful
to discriminate and to rule out various classes of inflationary models below the Planck scale.

A Gaussian random magnetic field, for a statistically
homogeneous and isotropic system, is described by the equal time two-point correlation function 
in momentum space as \cite{Shiraishi:2012rm,Finelli:2008xh}:
\be\begin{array}{lll}\label{eq1}
    \displaystyle \langle B^{*}_{i}({\bf k},\eta)B_{j}({\bf k^{'}},\eta)\rangle=(2\pi)^{3}\delta^{(3)}({\bf k}-{\bf k^{'}})
{\cal P}_{ij}(\hat{{\bf k}})
P_{\bf B}(k),
   \end{array}\ee

where the plane projector onto the transverse
plane is defined as \cite{Shiraishi:2012rm,Finelli:2008xh}:
\be\begin{array}{lllll}\label{eq2}
   \displaystyle {\cal P}_{ij}(\hat{{\bf k}})=\sum_{\lambda=\pm 1}e^{\lambda}_{i}(\hat{{\bf k}})
e^{-\lambda}_{j}(\hat{{\bf k}})=(\delta_{ij}-\hat{\bf k}_{i}\hat{\bf k}_{j})
   \end{array}\ee
in which the divergence-free nature of the magnetic field is imposed via the orthogonality condition, $\hat{\bf k}^{i}\epsilon^{\pm 1}_{i}=0$.
Here $\hat{\bf k}_{i}$ signifies the unit vector which can be expanded in terms of spin spherical harmonics. See \cite{Shiraishi:2012rm} for details.
Additionally, it is worthwhile to mention that in the present context,
$P_{\bf B}(k)$ be the   
part of the power spectrum
for the primordial magnetic
field which will only contribute to the cosmological
perturbations for the scalar modes and the Faraday Rotation at the phase of decoupling \cite{Shiraishi:2012rm,Shiraishi:2012sn}.

%%%%%%%%%%%%%%%%%%FIGURE%%%%%%%%%%%%%%%%%%%%%%%%%%%%%%%%%

%%%%%%%%%%%%%%%%%%%%%%%%%%%%%%%%%%%%%%%%%%%%%%%%%%%%%%%%%%%%%%%%%%%%%%%%%%%%%%%
%%%%%%%%%%%%%%%%%%%%%%%%%%%%%%%%%%%%%%%%%%%%%%%%%%%%%%%%%%%%%%%%%%%%%%%%%%%%%%%%%

The non-helical part of the primordial magnetic power spectrum is parameterized within the upper cut-off ($k_{\Lambda}$) and lower cut-off momentum scale ($k_{L}$) as:
\be\begin{array}{llll}\label{eq3}
\displaystyle  P_{\bf B}(k)=A_{\bf B}~\left(\frac{k}{k_{*}}\right)^{n_{\bf B}
+\frac{\alpha_{\bf B}}{2}\ln \left(\frac{k}{k_{*}}\right) +\frac{\kappa_{\bf B}}{6}\ln^{2} \left(\frac{k}{k_{*}}\right)+\cdots} 
\end{array}         
\ee
where $A_{\bf B}$ represents amplitude of the magnetic power spectrum,
 $n_{\bf B}$ is the spectral tilt, $\alpha_{\bf B}$ is the running and $\kappa_{\bf B}$ be the running of the spectral tilt. Here the upper
 cut-off momentum scale ($k_{\Lambda}$) corresponds
to the Alfv$\acute{e}$n wave damping length-scale, representing the dissipation of magnetic energy due to
the generation of magneto-hydrodynamic (MHD) waves. Also the lower cut-off momentum scale ($k_{L}$) corresponds to momentum scale corresponding present epoch $k_{0}$ 
which can be usually determined from the lower value of the multipole $l\sim2$
in the CMB angular power spectrum. Additionally, $k_{*}$ being the pivot or normalization scale of momentum.
Here the scale invariant magnetic power spectrum can be obtained from Eq~(\ref{eq3}) in the limit $n_{\bf B}\rightarrow 0$,
 $\alpha_{\bf B}\rightarrow 0$ and $\kappa_{\bf B}\rightarrow 0$.
But the most recent observational constraint
from CMB temperature anisotropies on the amplitude
and the spectral index of a primordial magnetic field has
been predicted by using Planck data as 
$B_{1~ Mpc} < 3.4 {\rm nG}$ prefers $n_{\bf B} < 0$ \cite{Ade:2013zuv} which show deviation from the scale invariance for the magnetic power spectrum.
Also if, in near future, Planck or any other observational probes can predict the signatures for running $\alpha_{\bf B}$ and running of the running $\kappa_{\bf B}$
 in the primordial
magnetic power spectrum (as already predicted in case of primordial scalar power spectrum within $1.5-2\sigma$ CL by Planck \cite{Ade:2013uln}
 and BICEP2 \cite{Ade:2014xna}), then it is possible to put 
further stringent constraint on
the various classes of inflationary models.

I am now interested in the mean square amplitude of the primordial magnetic field on a given characteristic scale $\xi$, on which I smooth the magnetic power
spectrum using a Gaussian filter as given by \cite{Finelli:2008xh}:
\be\begin{array}{lll}\label{eq4}
    \displaystyle B^{2}_{\xi}=\langle B_{i}({\bf x})B_{i}({\bf x})\rangle_{\xi}\\
\displaystyle~~~~
=\frac{1}{2\pi^{2}}\int^{\infty}_{0}\frac{dk}{k_{*}}~\left(\frac{k}{k_{*}}\right)^{2}P_{\bf B}(k)\exp\left(-k^{2}\xi^{2}\right).
   \end{array}\ee
Here Eq~(\ref{eq3}) describes a more generic picture where 
the magnetic power spectrum deviates from its exact power law form in presence of logarithmic correction. Consequently, the resulting 
mean square primordial magnetic field is logarithmically divergent in both the limits of the integral as presented in Eq~(\ref{eq4}).
To remove the divergent contribution from the mean square amplitude of the primordial magnetic field, 
I introduce here cut-off regularization technique in which I have re-parameterized the integral in terms of
regulated UV (high) and IR (low) momentum scales. Most importantly the
cut-offs $k_{\Lambda}$ and $k_{L}$ are momentum regulators to collect
only the finite contributions from Eq~(\ref{eq4}). Finally I get:
\begin{widetext}
\be\begin{array}{lll}\label{eq5}
    \displaystyle B^{2}_{\xi}(k_{L};k_{\Lambda})%\\ \displaystyle
 =\frac{A_{\bf B}}{2\pi^{2}}\int^{k_{\Lambda}}_{k_{L}}\frac{dk}{k_{*}}~\exp\left(-k^{2}\xi^{2}\right)%\\
%~~~~~~~~~~~~~~~~~~~~~~~~~\displaystyle\times
\left(\frac{k}{k_{*}}\right)^{n_{\bf B}+2
+\frac{\alpha_{\bf B}}{2}\ln \left(\frac{k}{k_{*}}\right) +\frac{\kappa_{\bf B}}{6}\ln^{2} \left(\frac{k}{k_{*}}\right)+\cdots}%\\
%,%\\
%\displaystyle~~~~
%~~~~~~~~~~~~~~
=\displaystyle\frac{I_{\xi}(k_{L};k_{\Lambda})}{2\pi^{2}}A_{\bf B}.
   \end{array}\ee
\end{widetext}
where the regularized integral function $I_{\xi}(k_{L};k_{\Lambda})$ is explicitly mentioned in the appendix.
On the other hand, in absence of any Gaussian filter, the magnetic energy density can be expressed in terms of the
 mean square primordial magnetic field as \cite{Finelli:2008xh}:
\be\begin{array}{lll}\label{eq6}
    \displaystyle \rho_{\bf B}=\frac{1}{8\pi}\langle B_{i}({\bf x})B_{i}({\bf x})\rangle%\\
%\displaystyle~~~~
=\frac{1}{8\pi^{2}}\int^{\infty}_{0}\frac{dk}{k_{*}}~\left(\frac{k}{k_{*}}\right)^{2}P_{\bf B}(k)
   \end{array}\ee
which is logarithmically divergent in UV and IR end. After introducing the momentum cut-offs I get the 
regularized expression for the magnetic energy density as:
\begin{widetext}
\be\begin{array}{lll}\label{eq7}
    \displaystyle \rho_{\bf B}(k_{L};k_{\Lambda})%\\ \displaystyle 
=\frac{A_{\bf B}}{8\pi^{2}}\int^{k_{\Lambda}}_{k_{L}}\frac{dk}{k_{*}}
\left(\frac{k}{k_{*}}\right)^{n_{\bf B}+2
+\frac{\alpha_{\bf B}}{2}\ln \left(\frac{k}{k_{*}}\right) +\frac{\kappa_{\bf B}}{6}\ln^{2} \left(\frac{k}{k_{*}}\right)+\cdots}%\\
%,%\\
%\displaystyle~~~~
\displaystyle =\frac{J(k_{L};k_{\Lambda})B^{2}_{\xi}(k_{L};k_{\Lambda})}{4I_{\xi}(k_{L};k_{\Lambda})}
   \end{array}\ee
\end{widetext}
where I use Eq~(\ref{eq5}). Here the regularized integral function $J(k_{L};k_{\Lambda})$ is explicitly written in the appendix.

Now to derive a phenomenological constraint here I further assume the fact that
 the primordial magnetic field is made up of relativistic species (bosons and fermions) which are originated from MSSM, MSSM$\otimes U(1)_{B_L}$ etc. particle theory.
In this physical prescription, the regularized magnetic energy
density can be expressed as \cite{Long:2013tha}:
\be\begin{array}{llll}\label{eq8}
    \displaystyle \rho_{\bf B}(k_{L};k_{\Lambda})\sim \frac{\pi^{2}}{30}g_{*}T^{4}\sim{\cal O}(10^{-13})
\times \frac{T^{4}}{\epsilon_{\bf CP}}
   \end{array}\ee
where $g_{*}$ be the degree of relativistic degrees of freedom, which can be expressed as:
 $g_{*}=g^{*}_{B}+\frac{7}{8}g^{*}_{F}$, where $g^{*}_{B}$ and $g^{*}_{F}$ characterize
the contributions from bosonic and fermionic degrees of freedom~\footnote{ For MSSM $g_{*}= 228.75$ considering all degrees of freedom.}.  
Also the CP asymmetry parameter is defined as:
 \be\begin{array}{lll}\label{eq9cvx}
     \displaystyle \epsilon_{\bf CP}=\frac{\Gamma_{L}(N_{R}\rightarrow L_{i}\Phi)-\Gamma_{L^c}(N_{R}\rightarrow L^{c}_{i}\Phi^{c})
}{\Gamma_{L}(N_{R}\rightarrow L_{i}\Phi)+\Gamma_{L^c}(N_{R}\rightarrow L^{c}_{i}\Phi^{c})}\lesssim {\cal O}(|\lambda|^2)
    \end{array}\ee
for the standard leptogenesis scenario \cite{Fukugita:1986hr,Buchmuller:2005eh} where the Majorana neutrino ($N_{R}$) decays through Yukawa matrix interaction ($\lambda$) with the 
Higgs ($\Phi$) and lepton ($L$) doublets. Here $\Gamma_{L}$ and $\Gamma_{L^c}$ are the corresponding decay widths for the process $N_{R}\rightarrow L_{i}\Phi$ and 
$N_{R}\rightarrow L^{c}_{i}\Phi^{c}$ respectively.
Now combining Eq~(\ref{eq7}) and Eq~(\ref{eq8}) I derive the following simplified expression for the root mean square
value of the primordial magnetic field at the present epoch in terms of the CP asymmetry parameter ($\epsilon_{\bf CP}$) as:
\be\begin{array}{lll}\label{eq11}
 \displaystyle B_{0}\sim
{\cal O}(10^{-14})\times\sqrt{\frac{I_{\xi}(k_{L}=k_{0};k_{\Lambda})}{J(k_{L}=k_{0};k_{\Lambda})\epsilon_{\bf CP}}}~{\rm Gauss}
\end{array}\ee
where I use the temperature at the present epoch $T_{0}\sim 2\times 10^{-4}$~eV and $1~{\rm Gauss}=7\times 10^{-20}$~${\rm GeV^{2}}$.
In addition, here I fix the momentum cut-off scale $k_{L}$ at the momentum scale corresponding to the 
present epoch $k_{0}\propto \eta^{-1}_{0}$, where $\eta_{0}\sim 14000$~{\rm Mpc} be the conformal time at the present epoch~\footnote{It is important to mention that once we put the lower IR cut-off scale,
 $k_{L}=k_{0}$ then the predicted value of the magnetic field from leptogenesis using Eq~(\ref{eq15}) and from inflation using Eq~(\ref{eq11}) is exactly same 
provided the inflation has to be embedded on a particle theory (Example: MSSM, MSSM$\otimes U(1)_{B-L}$ etc.) where the VEV and field excursion of the inflaton
generated from gauge invariant (supersymmetric) flat directions are sub-Planckian.}. Consequently the momentum integrals satisfy the following constraint:
\be \sqrt{\frac{I_{\xi}(k_{L}=k_{0};k_{\Lambda})}{J(k_{L}=k_{0};k_{\Lambda})}}\sim 10^{-8}.\ee

The conformal symmetry of the quantized electromagnetic field breaks down in curved space-time which 
is able to generate a sizable amount of magnetic field during a phase
of slow-roll inflation. Such primordial magnetism is characterized by the renormalized mean square amplitude
 of the primordial magnetic field at leading order in slow-roll approximation for comoving observers as \cite{Agullo:2013tba}:
\be\begin{array}{llll}\label{eq12}
    \displaystyle  \rho_{\bf B}(k_{L};k_{\Lambda})=\frac{1}{8\pi}\langle B_{i}({\bf x})B_{i}({\bf x})\rangle \approx\frac{V^{2}\epsilon_{V}}{2160\pi^{3}M^{4}_{p}}
   \end{array}\ee

 where $\epsilon_{V}=(M_{p}V^{'}/\sqrt{2}V)^2$ is the standard potential dependent slow-roll parameter
and $M_{p}\sim 2.43 \times 10^{18}~{\rm GeV}$ be the reduced Planck mass. Also here the
generic inflationary potential, $V(\phi)$ can be expanded around the vicinity of
 sub-Planckian VEV of inflation, $\phi_0(<M_{p})$ in a Taylor series as: 
\be\label{eqxc} V(\phi)=\alpha+\beta(\phi-\phi_{0})+\delta(\phi-\phi_{0})^{2}+\gamma(\phi-\phi_{0})^{3}+\kappa(\phi-\phi_{0})^{4}+\cdots\,\ee
where $\alpha$ denotes the height of the potential which sets the
 scale of inflation, and the coefficients $\beta,~\delta~\gamma,~\kappa$ determine the shape of the 
potential in terms of the model parameters for which the model is fully
embedded within a particle theory such as that of gauge
invariant flat directions of Minimal Supersymmetric Standard Model (MSSM), or MSSM$\otimes U(1)_{B-L}$.
Very recently in \cite{Choudhury:2014kma,Choudhury:2014wsa} I have investigated the detailed features of
 reconstruction technique of the generic inflationary potential for sub Planckian VEV and field excursion, which suggests that it is possible to generate 
the tensor-to-scalar ratio as observed by BICEP2 \cite{Ade:2014xna}, provided feasible amount of running and running of the running has to appear
 in the inflationary power spectrum both for scalar and tensor modes.
For further numerical estimations and checking the validity of the analysis prescribed in this paper I use the generic
 inflection point models
of inflation in which an additional
flatness condition, $\delta=V^{\prime\prime}(\phi_0)/2\approx 0$ is imposed in the generic expansion of the potential mention in Eq~(\ref{eqxc}) \cite{Choudhury:2013iaa,Choudhury:2013woa,Choudhury:2013jya}. 
Low scale MSSM inflation, high scale supergravity induced inflation are the well known examples of inflection point models which can 
be able to generate the tensor-to-scalar ratio as observed by Planck and BICEP2 data \cite{Choudhury:2011jt,Choudhury:2013jya,Choudhury:2014kma}.
  However this technique can be applicible to any models of inflation where the VEV is lying below the Planck scale.
It is important to note that Eq~(\ref{eq12}) is insensitive to
the intrinsic ambiguities of renormalization in curved space-times.
In general, one can expand $\epsilon_{V}$ at the momentum scale, $k_{L}\leq k\leq k_{\Lambda}$ around the pivot/normalization scale $k_{*}(\sim 0.002~{\rm Mpc}^{-1})$ as:
\be\begin{array}{llll}\label{eq13}
     \epsilon_{V}(k)=\epsilon_{V}(k_{*})-\frac{\alpha_{T}(k_{*})}{2}\ln\left(\frac{k}{k_{*}}\right)+\frac{\kappa_{T}(k_{*})}{4}\ln^{2}\left(\frac{k}{k_{*}}\right)+\cdots
  \end{array}\ee
where $\alpha_{T}(k_{*})$ and $\kappa_{T}(k_{*})$ are the running and
 running of running of the tensor spectral tilt evaluated at pivot $k=k_{*}$ scale and contributes in the next to leading order of
 the effective theory below UV cut-off. See the appendix where I have mentioned the inflationary consistency conditions which will contribute in Eq~(\ref{eq13}).

Let me fix the momentum scale at the pivot $k=k_{*}$ at which I can write,
 \be \epsilon_{V}(k_{*})\approx \frac{r(k_{*})}{16}
\left[1-2{\cal C}_{E}\left(\eta_{V}(k_{*})+\frac{r(k_{*})}{8}\right)\right]+\cdots,\ee where $\cdots$ includes the all the higher
 order slow-roll contributions. Here $\eta_{V}=(M^{2}_{p}V^{''}/V)$ be the another slow-roll parameter and ${\cal C}_{E}=4(\ln 2+\gamma_{E})-5$ with $\gamma_{E}=0.5772$ is the {\it Euler-Mascheroni constant}.
The recent observations from {\it Planck} puts an upper bound on  the amplitude of {\it primordial gravitational waves}
via tensor-to-scalar ratio, $r(k_*)=P_T/P_S$. This bounds the potential energy stored in the inflationary potential
~\cite{Ade:2013uln,Choudhury:2013iaa,Choudhury:2013woa,Choudhury:2013jya,Choudhury:2013qza}, i.e. 
\be\begin{array}{llll}\label{iopk}\displaystyle V_{*}\leq (1.96\times 10^{16}{\rm GeV})^{4}\left(\frac{r(k_{*})}{0.12}\right).\end{array}\ee 
 Finally using this constraint along with Eq~(\ref{eq7}) in Eq~(\ref{eq12}) I get the following simplified expression
for the root mean square value of
the primordial magnetic field in terms of the
tensor-to-scalar ratio $r$ and slow-roll parameter $\eta_{V}$ (defined in appendix) as:
\begin{widetext}
\be\begin{array}{llll}\label{eq15}
     \displaystyle B_{\xi}(k_{L};k_{\Lambda})\lesssim
{\cal O}(10^{44})\times\left(\frac{r(k_{*})}{0.12}\right)^{3/2}\Sigma^{3/2}(k_{L},k_{*})\sqrt{\frac{I_{\xi}(k_{L};k_{\Lambda})}{J(k_{L};k_{\Lambda})}\left[
1-2{\cal C}_{E}\left\{\eta_{V}(k_{*})+\frac{3}{200}\left(\frac{r(k_{*})}{0.12}\right)\right\}
+\cdots\right]}~{\rm Gauss}
    \end{array}\ee
\end{widetext}
where ${\cal C}_{E}=4(\ln 2+\gamma_{E})-5$ with $\gamma_{E}=0.5772$ is the {\it Euler-Mascheroni constant}.
At the present epoch the numerical factor $\Sigma(k_{L}=k_{0},k_{*})$ appearing in Eq~(\ref{eq15}) is lying within the window,
 $10^{-2/3}\leq \Sigma(k_{L}=k_{0},k_{*})\leq 10^{-30}$, for the tensor-to-scalar ratio, $10^{-29}\leq r_*\leq 0.12$ at the momentum pivot scale, 
$k_*\sim 0.002~{\rm Mpc}^{-1}$. 
Now by setting $k_{L}=k_{0}$ at the present epoch and using Eq~(\ref{eq11})
 we get the following expression for the lower bound of the CP asymmetry parameter
for generic sub-Planckian models of inflation as:
 \begin{widetext}
\be\begin{array}{llll}\label{eq16}
     \displaystyle \epsilon_{\bf CP}\gtrsim
{\cal O}(10^{-116})\times\left(\frac{0.12}{r(k_{*})}\right)^{3}\Sigma^{-3}(k_{L}=k_0,k_{*})\left[
1+2{\cal C}_{E}\left\{\eta_{V}(k_{*})+\frac{3}{200}\left(\frac{r(k_{*})}{0.12}\right)\right\}\right],
    \end{array}\ee
\end{widetext}
which is pointing towards the following possibilities:
$\Rightarrow${\bf (1)} For the small tensor-to-scalar ratio the significant features of CP asymmetry can be possible to detect 
in colliders experiments, 
$\Rightarrow${\bf (2)} For large value of tensor-to-scalar ratio the CP asymmetry is largely suppressed and can't be possible to detect 
in the particle colliders.
If, in near future, any direct/indirect observational probe detects the signatures of primordial gravitational waves by measuring large detectable amount of 
tensor-to-scalar ratio then it will follow the second possibility. On the other hand, in the first case it is highly
possible to achieve the upper bound of CP asymmetry parameter, $\epsilon_{\bf CP}\leq 10^{-6}$ 
by fixing the lower bound of the tensor-to-scalar ratio at very small value, $r(k_{*})\sim 10^{-29}$ at the pivot scale $k_{*}\sim 0.002~{\rm Mpc}^{-1}$
to accommodate Majorana neutrino at the scale of $10^{10}~{\rm GeV}$. 

Finally my prime objective is to study how the derived relations in our paper further put a stringent constraint on the CMB BB polarization power spectra.
To achieve this goal I start with
a comoving radiation frequency $\nu_{0}$ in which I consider the Faraday rotation angle $\alpha(\hat{\bf n})$ of the CMB linear
polarization as a function of sky direction $\hat{\bf n}$ along with the restriction that the full sky average of the rotation angle
is zero. Now expanding the Faraday rotation angle in terms of spherical harmonics which leads to the definition of the
angular power spectrum $C^{{\bf \alpha}}_{l}$ via the two-point correlation \cite{Kahniashvili:2008hx,Kosowsky:2004zh}:
\be\begin{array}{llll}\label{eq18xcz}
    \displaystyle \langle \alpha(\hat{\bf n})\alpha(\hat{\bf n^{'}})\rangle=\sum_{l}\frac{2l+1}{4\pi}C^{{\bf \alpha}}_{l}P_{l}(\hat{\bf n}\odot\hat{\bf n^{'}}),
   \end{array}\ee
where $P_{l}(\hat{\bf n}\odot\hat{\bf n^{'}})$ be the Legendre polynomial of order {\it l} and I have:
\begin{widetext}
\be\begin{array}{llll}\label{eq19}
    \displaystyle C^{{\bf \alpha}}_{l}\simeq\frac{9l(l+1)A_{\bf B}}{128\pi^{5}\alpha_{\bf EM}\nu^{4}_{0}}\int^{k_{\Lambda}}_{k_{L}=k_{0}}
\frac{dk}{k_{*}}~\left(\frac{k}{k_{*}}\right)^{n_{\bf B}+2
+\frac{\alpha_{\bf B}}{2}\ln \left(\frac{k}{k_{*}}\right)
 +\frac{\kappa_{\bf B}}{6}\ln^{2} \left(\frac{k}{k_{*}}\right)+\cdots}\left(\frac{{\cal J}_{l}(k\eta_{0})}{k\eta_{0}}\right)^{2}
=\frac{9l(l+1)B^{2}_{0}{\cal M}(k_{0},k_{\Lambda})}{64\pi^{3}\alpha_{\bf EM}\nu^{4}_{0}I_{\xi}(k_{0};k_{\Lambda})}
   \end{array}\ee
\end{widetext}
where ${\cal M}(k_{0},k_{\Lambda})$ be the corresponding integral kernel, $\alpha_{\bf EM}\approx 1/137$ be the fine structure constant,
$\nu_{0}$ be the comoving frequency of the observed radiation and ${\cal J}_{l}(k\eta_{0})$ be the Spherical Bessel function of order ``l''.
Additionally, it is important to mention that in order to compute the momentum integral appearing in Eq~(\ref{eq19}) I first substitute Eq~(\ref{eq15}) and fix the lower cut-off of the 
momentum scale at $k_{L}=k_{0}$. However Eq~(\ref{eq15}) clearly suggests that the contribution from the primordial tensor modes via tensor to scalar ratio from 
inflation is considered within the window $10^{-29}<r_{*}<0.12$ which is only possible in 
 previously mentioned particle physics phenomenological setup. Within this background at $k_{L}=k_{0}$ both Eq~(\ref{eq11}) and Eq~(\ref{eq15}) carry the
equivalent physical information. Additionally contributions from the E-mode by Faraday rotation are also taken care of in this computation. 
Finally, using Eq~(\ref{eq19}) the B-mode polarization angular power spectrum induced by
the Faraday rotation field from the primordial E-mode polarization can be expressed as: 
\begin{widetext}
 \be\begin{array}{lll}\label{eq20}
 \displaystyle C^{BB}_{l}=\frac{9B^{2}_{0}{\cal M}(k_{0},k_{\Lambda})l(l+1)(l-2)!}{256\pi^{4}\alpha_{\bf EM}\nu^{4}_{0}I_{\xi}(k_{0};k_{\Lambda})(2l+1)(l+2)!}\sum_{l_{1},l_{2}}\left\{\frac{l_{1}(l_{1}+1)(2l_{1}+1)(2l_{2}+1)(l_{2}-2)!}{(l_{2}+2)!}
\left[l^{2}(l+1)^{2}+l^{2}_{1}(l_{1}+1)^{2}\right.\right.\\ \left.\left.~~~~~~~~~~~~~~~~~~~~~~~~~~~~~~~~~~~~~~~~~~~~~~~~~~~~~~~~~~~~~~~~~~~
\displaystyle +l^{2}_{2}(l_{2}+1)^{2}-2l_{1}l_{2}(l_{1}+1)(l_{2}+1)-2l_{1}l(l_{1}+1)(l+1)\right.\right.\\ \left.\left.
~~~~~~~~~~~~~~~~~~~~~~~~~~~~~~~~~~~~~~~~~~~~~~~~~~~~~~~~~~~~~~~~~~~~~~+2\left\{l_{1}(l_{1}+1)-l_{2}(l_{2}+1)-l(l+1)\right\}
\right]C^{EE}_{l_{2}}\left({\bf \Delta}^{l0}_{l_{1}0;l_{2}0}\right)^{2}\right\}
    \end{array}\ee

\end{widetext}

where the regularized angular power spectrum for the primordial E-mode polarization can be expressed as:
%\\
\begin{widetext}
\be\begin{array}{lll}\label{eq21}
    \displaystyle C^{EE}_{l_{2}}=16\pi^{2}P_{\bf S}(k_{*})\int^{k_{\Lambda}}_{k_{L}=k_{0}}\frac{dk}{k_{*}}~\Theta^{2}_{El_{2}}(k)
\left(\frac{k}{k_{*}}\right)^{n_{\bf S}(k_*)+1
+\frac{\alpha_{\bf S}(k_*)}{2}\ln \left(\frac{k}{k_{*}}\right) +\frac{\kappa_{\bf S}(k_*)}{6}\ln^{2} \left(\frac{k}{k_{*}}\right)+\cdots}
=16\pi^{2}P_{\bf S}(k_{*})\vartheta(k_{0},k_{\Lambda}).
   \end{array}
\ee
\end{widetext}
%%%%%%%%%%%%%%%%%%FIGURE%%%%%%%%%%%%%%%%%%%%%%%%%%%%%%%%%

\begin{figure}[htb]
\centering
\includegraphics[width=8.5cm,height=8.5cm]{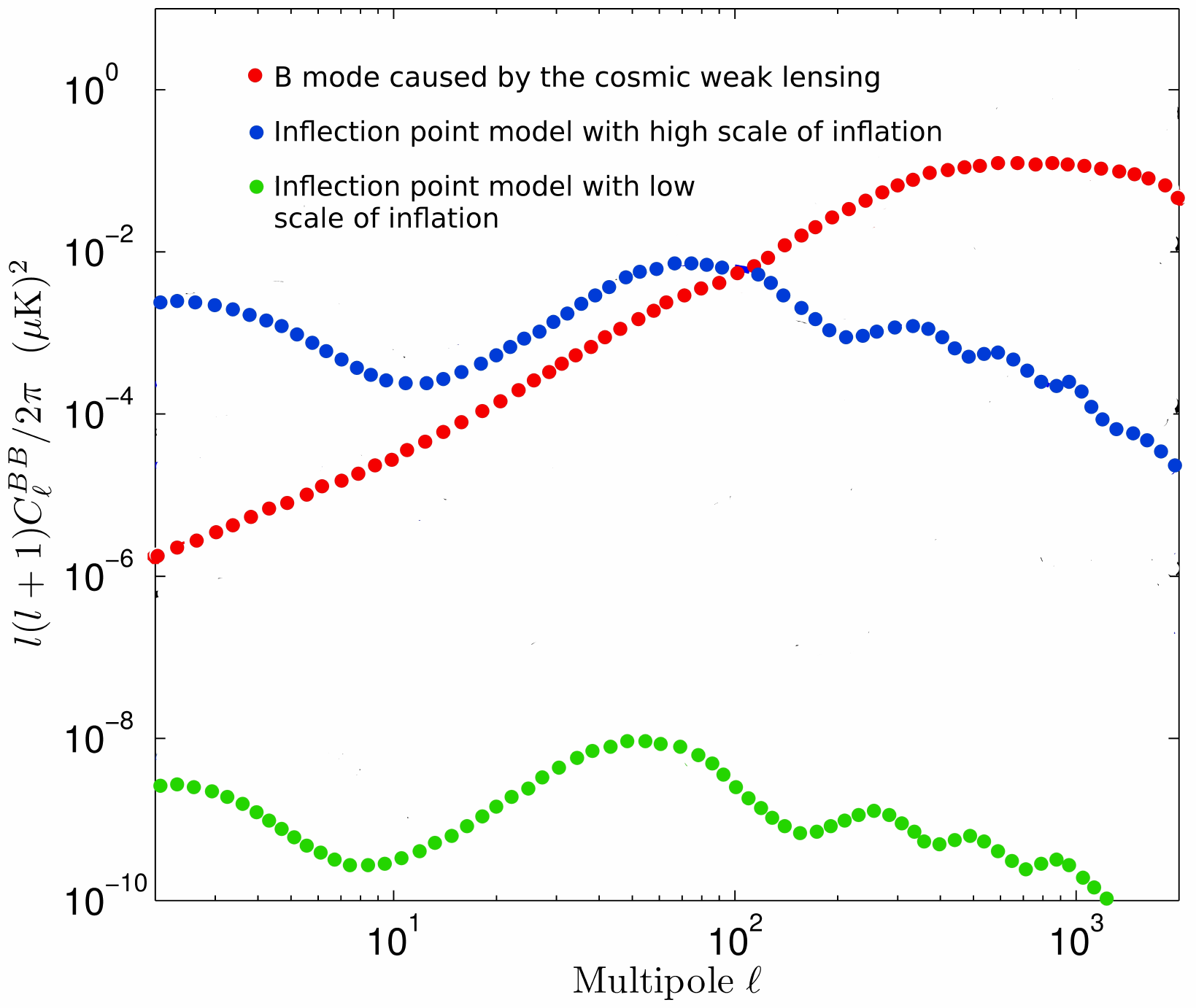}
\caption{Variation of CMB B mode angular power spectra ($l(l+1)C^{BB}_{l}/2\pi$) with respect to multipole ($l$). In this analysis the Faraday rotation angle assumed to
be maximal everywhere on the sky.
The blue and green dotted curves correspond to the sub-Planckian inflection point models of inflation with low scale, $V^{1/4}_*=6.48\times 10^{8}$GeV (with $r_{*}\sim10^{-29}$) and high scale
,$V^{1/4}_*=1.96\times 10^{16}$GeV (with $r_{*}\sim 0.12$) at momentum pivot scale $k_{*}=0.002~{\rm Mpc}^{-1}$. 
For both blue and green curves, the numerical value of the magnetic field at the present epoch is, $B_{0}\sim 10^{-9}$~Gauss.
But for them the CP asymmetry parameter is different, $\epsilon_{\bf CP}\gtrsim 10^{-26}$ (for blue curve) and $\epsilon_{\bf CP}\gtrsim10^{-30}$ (for green curve).
The red dotted curve corresponds to the B mode caused due to the
cosmic weak lensing. The bounded region between the blue and green dotted curves are the
 theoretically allowed region of generic inflection point inflationary models. For an example, within the framework of MSSM 
the inflection point technique holds good for the sub-Planckian VEV, $\phi_{0}$. 
}
\label{f1}
\end{figure}
%%%%%%%%%%%%%%%%%%%%%%%%%%%%%%%%%%%%%%%%%%%%%%%%%%%%%%%%%%%%%%%%%%%%%%%%%%%%%%%
%%%%%%%%%%%%%%%%%%%%%%%%%%%%%%%%%%%%%%%%%%%%%%%%%%%%%%%%%%%%%%%%%%%%%%%%%%%%%%%%%
where $\vartheta(k_{0},k_{\Lambda})$ be the corresponding integral kernel.
Here the subscript ${\bf S}$ represents the scalar modes and the momentum dependent integral function \cite{Baumann:2009ds},
 \be \Theta_{El_{2}}(k)=\int^{\eta_{0}}_{0}d\eta~K_{E}(k,\eta){\cal P}_{El_{2}}(k(\eta_{0}-\eta)),\ee where $K_{E}(k,\eta)$ characterizes the physical source kernel
and ${\cal P}_{El_{2}}(k(\eta_{0}-\eta))$ represents the geometric projection factor
which can be expressed in terms of the spherical Bessel functions. 
Also ${\bf \Delta}^{l0}_{l_{1}0;l_{2}0}$ are Clebsch-Gordan coefficients can be written in a closed form as \cite{book}:
\be\begin{array}{llll}\label{eq22}
    \displaystyle {\bf \Delta}^{l0}_{l_{1}0;l_{2}0}=\frac{2(-1)^{n-l}n!\sqrt{2(2l+1)}}{\sqrt{(n-l_{1})!(n-l_{2})!(n-l)!(2n+1)!}}
   \end{array}
\ee
which is valid for $l_{1}+l_{2}+l=2n$, with $n$ be a positive integer; $l_{1}$, $l_{2}$ and $l$ must also satisfy the triangle inequalities.
In the present context the Clebsch Gordan coefficients can be directly evaluated numerically, as long as the various factors are taken in an order to prevent
overflow and/or underflow errors. Inspite of that the direct computation of these coefficients is not fast enough because 
it involves several thousands of Clebsch-Gordan coefficients during the summation over the multipoles $l$.
In Eq~(\ref{eq20}) the B-mode polarization angular power spectrum is constrained as 
the magnetic field at the present epoch $B_{0}$ satisfies all the previously derived constraints in terms of tensor-to-scalar ratio ($r$) and 
CP asymmetry parameter ($\epsilon_{\bf CP}$). The modifications to the existing E-polarization power spectrum and the
cross-correlation between E-polarization and temperature are negligible for small rotation angles, and consequently I ignore the contribution from the 
second-order effects as they are sufficiently small in the prescribed analysis presented in this paper.

In Fig~(\ref{f1}) I have plotted the behaviour of the CMB B mode angular power spectra from BB correlation with respect to the multipole.
Using Eq~(\ref{eq20}) I have obtained the blue and green dotted theoretical curves which
 correspond to the sub-Planckian inflection point models of inflation with low scale, $V^{1/4}_*=6.48\times 10^{8}$GeV (with $r_{*}\sim10^{-29}$) 
and high scale, $V^{1/4}_*=1.96\times 10^{16}$GeV (with $r_{*}\sim 0.12$) at pivot scale of momentum, $k_{*}=0.002~{\rm Mpc}^{-1}$. 
For both blue and green curves, the numerical value of the magnetic field at the present epoch is, $B_{0}\sim 10^{-9}$~Gauss.
But for them the CP asymmetry parameter is different, $\epsilon_{\bf CP}\gtrsim 10^{-26}$ (for blue curve) and $\epsilon_{\bf CP}\gtrsim10^{-30}$ (for green curve).
The red dotted curve corresponds to the B mode caused due to the
cosmic weak lensing. Here the blue curve also signifies the present Planck $2\sigma$ constraint on the upper bound of tensor to scalar ratio at $r_{*}\leq 0.12$.
Within the allowed window of tensor-to-scalar ratio for inflection point
 inflationary models, $10^{-29}\leq r_{0.002}\leq 0.12$, the amplitude of scalar power spectrum $(P_{S}(k_*))$, spectral tilt ($n_{S}(k_*)$), running ($\alpha_{S}(k_*)$)
and running of the running ($\kappa_{S}(k_*)$) lying within:
$2.092<10^{9}P_{S}(k_*)<2.297\,$, $0.958<n_{S}(k_*)<0.963\,$,
$-0.0098<\alpha_{S}(k_*)<0.0003\,$ and 
$-0.0007<\kappa_{S}(k_*)<0.006\,$.
which confronts the {\it Planck}+WMAP-9+BAO data set, \cite{Ade:2013uln,Ade:2013zuv}, well within $2\sigma$ CL.
For an example, within the framework of MSSM, where soft supersymmetry breaking sector (in low scale) \cite{Wang:2013hva} 
and Hubble induced supergravity non-minimal K\"ahler correction \cite{Choudhury:2013jya,Choudhury:2014sxa,Choudhury:2014uxa,Choudhury:2013zna}
computed from string moduli based hidden sector \cite{Choudhury:2011sq,Choudhury:2011rz,Choudhury:2012ib,Choudhury:2012yh}
(in high scale) plays a pivotal role in particle physics, the proposed technique holds good for the sub-Planckian
 VEV, $\phi_{0}<M_{p}$ and field excursion $\Delta\phi<M_{p}$. 
Most importantly, once Planck or any other observational probe detect the existence of primordial gravitational waves in near future 
then it possible to comment on the scale of inflation using which one can further tightly constrain the low or high
scale inflection point inflationary models below the Planck scale, $M_{p}$~\footnote{BICEP2 confirms the existence of primordial gravity waves at the multipole 
$l\sim 80$ via tensor-to-scalar
ratio within the window, $0.15<r_{*}<0.27$ ($r_{*}=0$ is ruled out by $7\sigma$ CL), which will constrain the scale of the potential within,
$2.07\times10^{16}~{\rm GeV}\leq\sqrt[4]{V_{\star}}\leq 2.40\times 10^{16}~{\rm GeV}$. After foreground subtraction the window of tensor-to-scalar ratio becomes,
$0.11<r_{*}<0.22$, which will further constrain the scale of inflation within the window,
 $1.91\times10^{16}~{\rm GeV}\leq\sqrt[4]{V_{\star}}\leq 2.28\times 10^{16}~{\rm GeV}$. This clearly suggests that high scale inflationary models are highly favoured
compared to the low scale models after releasing the BICEP2 data.}.

To summarize, in the present article, I have
established a generic connecting relationship
 between inflationary magnetogensis and primordial gravitational waves via tensor-to-scalar ratio and CP asymmetry parameter 
for a generic sub-Planckian model of inflation with a flat potential, 
 where inflation is driven near an inflection-point. 
For such a class of model it is also possible to predict amount of magnetic field at the present epoch
by measuring CP asymmetry parameter or the tensor-to-scalar ratio. 
Most significantly, once the signature of primordial gravity waves will be predicted, it will be possible to comment on the associated CP asymmetry and vice versa.
I have used important constraints arising from Planck on amplitude of power spectrum, spectral tilt and its running within $1.5\sigma-2\sigma$ statistical CL. To this end 
I have shown the behaviour of theoretical CMB B mode polarization power spectra for low and high scale inflection point models of inflation within
$10^{-29}\leq r_{0.002} \leq 0.12$, which is 
also consistent with the upper bound of tensor-to-scalar ratio obtained from Planck data. Further my aim is to carry forward this work 
in a more broader sense where I will extract the contribution from gravitational lensing and non-Gaussian contribution from the primordial magnetic field
from the generic sub-Planckian inflationary setup expected to be reported shortly \cite{sayan}.

%%%%%%%%%%%%%%%%%%%%%%%%%%%%%%%%%%%%%%%%%%%%%%%%%%%%%%%%%%%%%%%%%%%%%%%%%%%%%%%%%%%%%%%%%%%%%%%%%%%%%%%%%%%%%%%%%%%%%%%%%%%%%%%%%%%%%%%%%%%%%%%%%%%%%%%%%%%%%%%%%%%%%%%%%%%%%%%%%%%%%%%%%%%%%%%%%%%%%%%%%%%%%%%%%%%%%%%%%%%%%
%%%%%%%%%%%%%%%%%%%%%%%%%%%%%%%%%%%%%%%%%%%%%%%%%%%%%%%%%%%%%%%%%%%%%%%%%%%%%%%%%%%%%%%%%%%%%%%%%%%%%%%%%%%%%%%%%%%%%%%%%%%%%%%%%%%%%%%%%%%%%%%%%%%%%%%%%%%%%%%%%%%%%%%%%%%%%%%%%%%%%%%%%%%%%%%%%%%%%%%%

\section*{Acknowledgments}

%{\bf Acknowledgments:}
 I would like to thank Council of Scientific and
Industrial Research, India for financial support through Senior
Research Fellowship (Grant No. 09/093(0132)/2010). 
 I take this opportunity to thank the organizers of 8th Asian School on Strings, Particles and Cosmology, 2014 for the hospitality during the
work. Last but not the least I would like to thank sincerely to Prof. Soumitra SenGupta, Prof. Anupam Mazumdar and Dr. Supratik Pal for their constant support and inspiration.
%%%%%%%%%%%%%%%%%%%%%%%%%%%%%%%%%%%%%%%%%%%%%%%%%%%%%%%%%%%%%%%%%%%%%%%%%%%%%%%%%%%%%%%%%%%%%%%%%%%%%%%%%%%%%%%%%%%%%%%%%%%%%%%%%%%%%%%%%%%%%%%%%%%%%%%%%%%%%%%%%%%%%%%%%%%%%%%%%%%%%%%%
%%%%%%%%%%%%%%%%%%%%%%%%%%%%%%%%%%%%%%%%%%%%%%%%%%%%%%%%%%%%%%%%%%%%%%%%%%%%%%%%%%%%%%%%%%%%%%%%%%%%%%%%%%%%%%%%%%%%%%%%%%%%%%%%%%%%%%%%%%%%%%%%%%%%%%%%%%%%%%%%%%%%%%%%%%%%%%%%%%%%%%%%%%%%%%%%%%%
\section*{Appendix}
%{\bf Appendix:} 
$~~~~~~~~~~~~~~~~$\underline{\bf A. Momentum Integrals:}\\ \\
Additionally the momentum integrals used in Eq~(\ref{eq5},\ref{eq7},\ref{eq11},\ref{eq15}) can be written as:
\begin{widetext}
 \be\begin{array}{llll}\label{eq17}
  \displaystyle I_{\xi}(k_{L},k_{\Lambda}):= \left[\frac{\sqrt{\pi}~Erf\left(\xi k\right)}{2\xi k_{*}}\left\{1+{\cal Q}\ln\left(\frac{k}{k_{*}}\right)
+{\cal P}\ln^{2}\left(\frac{k}{k_{*}}\right)+{\cal F}\ln^{3}\left(\frac{k}{k_{*}}\right)\right\}\right.\\ \left. 
\displaystyle~~~~~~~~~~~~~~~~~~~~~~~~~~~~~~~~~~~~~~~~~~~~~~~~~ +\left(\frac{k}{k_{*}}\right)
\left\{-6{\cal F}~_PF_Q\left[\left\{\frac{1}{2},\frac{1}{2},\frac{1}{2},\frac{1}{2}\right\}~;\left\{\frac{3}{2},\frac{3}{2},\frac{3}{2},\frac{3}{2}\right\}~;
-\xi^{2}k^{2}\right]\right.\right.\\ \left.\left.\displaystyle 
~~~~~~~~~~~~~~~~~~~~~~~~~~~~~~~~~~~~~~~~~~~~~~~~~~+2\left({\cal P}+3{\cal F}\ln\left(\frac{k}{k_{*}}\right)\right)
~_PF_Q\left[\left\{\frac{1}{2},\frac{1}{2},\frac{1}{2}\right\}~;\left\{\frac{3}{2},\frac{3}{2},\frac{3}{2}\right\}~;
-\xi^{2}k^{2}\right]\right.\right.\\ \left.\left.\displaystyle 
~~~~~~~~~~~~~~~~~~~~~~~~~~~~~~~~~~~~~~~~~~~~~~~~~~
\displaystyle -\left({\cal Q}+2{\cal P}\ln\left(\frac{k}{k_{*}}\right)+6{\cal F}\ln^{2}\left(\frac{k}{k_{*}}\right)\right)
~_PF_Q\left[\left\{\frac{1}{2},\frac{1}{2}\right\}~;\left\{\frac{3}{2},\frac{3}{2}\right\}~;
-\xi^{2}k^{2}\right]\right\}\right]^{k=k_{\Lambda}}_{k=k_{L}},\\
 \displaystyle J(k_{L},k_{\Lambda}):= \left(\frac{k}{k_{*}}\right)\left[(1-6{\cal F}+2{\cal P}-{\cal Q})+(6{\cal F}-2{\cal P}+{\cal Q})\ln\left(\frac{k}{k_{*}}\right)
-(3{\cal F}-{\cal P})\ln^{2}\left(\frac{k}{k_{*}}\right)+{\cal F}\ln^{3}\left(\frac{k}{k_{*}}\right)\right]^{k=k_{\Lambda}}_{k=k_{L}}.
    \end{array}\ee
\end{widetext}
where ${\cal Q}=n_{\bf B}+2$, ${\cal P}=\alpha_{\bf B}/2$ and ${\cal F}=\kappa_{\bf B}/6$.\\ \\ \\ \\

$~~~~$\underline{\bf B. Inflationary Consistency Conditions:}\\ \\
In this paper we use the following slow-roll consistency relations at the pivot scale $k_{*}$: 
\begin{widetext}
\be\begin{array}{lll}\label{eq14}
    \displaystyle\epsilon_{V}(k_{*})\approx \frac{r(k_{*})}{16}\left[1-2{\cal C}_{E}\left(\eta_{V}(k_{*})+\frac{r(k_{*})}{8}\right)\right]+\cdots,\\
\displaystyle
\alpha_{T}(k_{*})\approx\frac{r(k_{*})}{8}\left[\frac{r(k_{*})}{8}+2\eta_{V}(k_{*})-\frac{3r(k_{*})}{8}\left\{1-2{\cal C}_{E}\left(\eta_{V}(k_{*})
+\frac{r(k_{*})}{8}\right)\right\}\right]+\cdots,\\
\displaystyle
\kappa_{T}(k_{*})\approx\frac{r(k_{*})}{8}\left[\frac{r(k_{*})}{4}-2\eta_{V}(k_{*})\right]\left[\frac{r(k_{*})}{4}+2\eta_{V}(k_{*})-\frac{3r(k_{*})}{8}\left\{1-2{\cal C}_{E}\left(\eta_{V}(k_{*})
+\frac{r(k_{*})}{8}\right)\right\}\right]+\cdots.
   \end{array}\ee
\end{widetext}

%%%%%%%%%%%%%%%%%%%%%%%%%%%%%%%%%%%%%%%%%%%%%%%%%%%%%%%%%%%%%%%%%%%%%%%%%%%%%%%%%%%%%%%%%%%%%%%%%%%%%%%%%%%%%%%%%%%%%%%%%%%%%%%%%%%%%%%%%%%%%%%%%%%%%%%%%%%%%%%%%%%%%%%%%%%%%%%%%%%%%%%%%%%
%%%%%%%%%%%%%%%%%%%%%%%%%%%%%%%%%%%%%%%%%%%%%%%%%%%%%%%%%%%%%%%%%%%%%%%%%%%%%%%%%%%%%%%%%%%%%%%%%%%%%%%%%%%%%%%%%%%%%%%%%%%%%%%%%%%%%%%%%%%%%%%%%%%%%%%%%%%%%%%%%%%%%%%%%%%

%\end{thebibliography}

\end{document}